\documentclass[conference]{IEEEtran}

\usepackage[dvips]{color}
\usepackage{graphicx}
\usepackage{amsmath}
\begin{document}
%
\title{Secrecy Outage Probability Analysis of Multi-User Multi-Eavesdropper Wireless Systems}

%
%

\author{
\IEEEauthorblockN{Yulong~Zou\IEEEauthorrefmark{1}, Jia~Zhu\IEEEauthorrefmark{1}, Gongpu~Wang\IEEEauthorrefmark{2}, and Hua~Shao\IEEEauthorrefmark{3}}

\IEEEauthorblockA{\IEEEauthorrefmark{1}School of Telecomm. \& Inform. Eng., Nanjing Univ. of Posts and Telecomm., Nanjing, China}
\IEEEauthorblockA{\IEEEauthorrefmark{2}School of Comput. and Inform. Tech., Beijing Jiaotong University, Beijing, China}
\IEEEauthorblockA{\IEEEauthorrefmark{3}China Mobile Research Institute, Beijing, China}
}

\maketitle

\begin{abstract}
In this paper, we explore the physical-layer security of a multi-user wireless system that consists of multiple users intending to transmit to a base station (BS), while multiple eavesdroppers attempt to tap the user transmissions. We examine the employment of multi-user scheduling for improving the transmission security against eavesdropping and propose a multi-user scheduling scheme, which only requires the channel state information (CSI) of BS without the need of the passive eavesdroppers' CSI. We also consider the round-robin scheduling for comparison purposes. The closed-form secrecy outage probability expressions of the round-robin scheduling and proposed multi-user scheduling are derived over Rayleigh fading channels. Numerical results demonstrate that the proposed multi-user scheduling outperforms the round-robin scheduling in terms of the secrecy outage probability. As the number of users increases, the secrecy outage probability of round-robin scheduling keeps unchanged. By contrast, the secrecy outage performance of the proposed multi-user scheduling improves significantly with an increasing number of users.

\end{abstract}


\IEEEpeerreviewmaketitle

\section{Introduction}
Recently, wireless security has attracted significant attention from the research community [1], [2]. The broadcast nature of radio propagation makes the wireless transmission accessible to unauthorized users and extremely vulnerable to the eavesdropping attack. Traditionally, cryptographic techniques relying on secret keys are employed to guarantee the transmission confidentiality. However, an eavesdropper with unlimited computing power can still crack the conventional cryptographic approaches by launching the exhaustive key search (known as brute-force attack). To this end, physical-layer security (PLS) is now emerging as a promising paradigm to protect the communication confidentiality against the brute-force attack by exploiting the physical characteristics of wireless channels [3].

PLS work can be traced back to 1970s and was studied by Wyner in [4], where a classical wiretap channel model comprised of one source and one destination in the presence of an eavesdropper was analyzed and characterized from an information-theoretic perspective in terms of the achievable secrecy rate performance. In [5], the authors introduced the notion of \emph{secrecy capacity}, which is shown as the difference between the channel capacity of the main link (from source to destination) and the wiretap link (from source to eavesdropper). However, in wireless networks, the multipath fading of radio propagation results in a performance degradation of the wireless secrecy capacity. As a remedy, the multiple-input multiple-output (MIMO) [6], [7] and cooperative relaying techniques [8], [9] were studied extensively to improve the wireless secrecy capacity against the fading effect.

The artificial noise aided security approaches [10]-[12] were examined by generating a specifically-designed interfering signal (referred to as artificial noise) that will adversely affect the wiretap channel only without causing any negative impact on the legitimate main channel. In [10], the authors proposed a quality of service (QoS) based artificial noise generation scheme for the sake of minimizing the signal-to-interference-and-noise ratio received at an eavesdropper while meeting a required signal-to-interference-plus-noise ratio (SINR) received at the desired destination. In [11], the multiple antennas assisted artificial noise generation was studied and the closed-form secrecy rate expressions were derived in high signal-to-noise ratio (SNR) regime under the target data throughput requirement. Additionally, a joint artificial noise and beamforming design was studied in [12] for enhancing the physical-layer security in the presence of multiple eavesdroppers.

It is worth mentioning that the security advantage of artificial noise methods studied in [10]-[12] comes at the expense of additional power resources which are consumed for generating the artificial noise to confuse the eavesdropper. By contrast, this paper examines the multi-user scheduling for improving the wireless physical-layer security without consuming any additional power resources. The main contributions of this paper can be summarized as follows. First, we propose a multi-user scheduling scheme without the channel state information (CSI) of eavesdroppers in a wireless network consisting of one base station (BS) and multiple users. The conventional round-robin scheduling is also considered as a benchmark. Second, we derive exact closed-form expressions of the secrecy outage probability for both the conventional round-robin scheduling and the proposed multi-user scheduling schemes over Rayleigh fading channels. Finally, numerical evaluations are carried out to show the security advantage of the proposed  multi-user scheduling over the round-robin scheduling in terms of the secrecy outage probability.

The remainder of this paper is organized as follows. In Section II, we describe the system model of a wireless network in the presence of multiple eavesdroppers. Then, we present the conventional round-robin scheduling as well as the proposed multi-user scheduling schemes in Section III, where the closed-form secrecy outage expressions of two scheduling schemes are derived over Rayleigh fading channels. Next, in Section IV, the numerical performance evaluation is conducted in terms of the secrecy outage probability. Finally, we provide some concluding remarks in Section V.

\section{System Model}
\begin{figure}
  \centering
  {\includegraphics[scale=0.65]{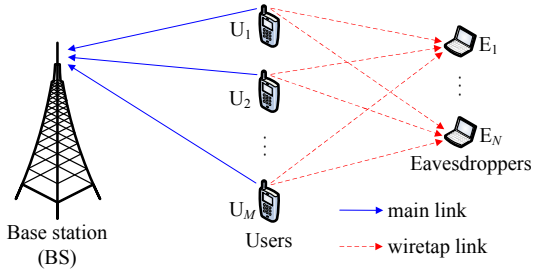}\\
  \caption{A multi-user multi-eavesdropper (MUME) wireless network.}\label{Fig1}}
\end{figure}
As illustrated in Fig. 1, we consider a multi-user multi-eavesdropper wireless network consisting of one base station (BS) and $M$ users in the presence of $N$ eavesdroppers. In Fig. 1, $M$ users are intended to transmit their data to the BS, while there are $N$ eavesdroppers attempting to tap the data transmitted from the users. For notational convenience, the set of $M$ users and that of $N$ eavesdroppers are denoted by ${\cal{U}}=\{{\textrm{U}}_i|i=1,2,\cdots,M\}$ and ${\cal{E}}=\{{\textrm{E}}_j|j=1,2,\cdots,N\}$, respectively. Without any loss of generality, we consider that $\textrm{U}_i$ transmits its signal $x_i$ to BS with power $P_i$, thus the signal received at BS can be expressed as
\begin{equation}\label{equa1}
{y_b} = \sqrt {P_i} {h_{ib}}{x_i} + {n_b},
\end{equation}
where ${h_{ib}}$ represents the wireless fading of the channel from $\textrm{U}_i$ to BS and $n_b$ represents the zero-mean additive white Gaussian noise (AWGN) with the variance of $N_0$ experienced at BS. From (1), we can obtain the channel capacity from $\textrm{U}_i$ to BS as
\begin{equation}\label{equa2}
{C_{ib}} = {\log _2}(1 +\frac{|{h_{ib}}{|^2}P_i}{{{N_0}}}).
\end{equation}
Meanwhile, due to the openness nature of wireless propagation, the signal $x_i$ transmitted from $\textrm{U}_i$ may also be overheard by the eavesdroppers. Hence, the received signal at an eavesdropper ${\textrm{E}}_j$ is expressed as
\begin{equation}\label{equa3}
{y_{e_j}} = \sqrt {P_i} {h_{ie_j}}{x_i} + {n_{e_j}},
\end{equation}
where ${h_{ie_j}}$ represents the wireless fading of the channel from $\textrm{U}_i$ to ${\textrm{E}}_j$ and $n_{e_j} $ represents the zero-mean AWGN with the variance of $N_0$ experienced at the eavesdropper ${\textrm{E}}_j$. Using (3), we obtain the channel capacity from $\textrm{U}_i$ to ${\textrm{E}}_j$ as
\begin{equation}\label{equa4}
{C_{ie_j}} = {\log _2}(1 +\frac{|{h_{ie_j}}{|^2}P_i}{{{N_0}}}).
\end{equation}

In this paper, the eavesdroppers of Fig. 1 are assumed to be uncoordinated in tapping the $\textrm{U}_i$-BS transmission. This implies that if any of the $N$ eavesdroppers succeeds in decoding the signal $x_i$, the $\textrm{U}_i$-BS transmission becomes insecure. Thus, the overall wiretap channel capacity is given by the maximum of individual channel capacities achieved at the $N$ eavesdroppers, yielding
\begin{equation}\label{equa5}
{C_{ie}}=\mathop {\max }\limits_{e_j \in {\cal{E}}} {C_{{ie_j}}} =  \mathop {\max }\limits_{e_j \in {\cal{E}}} {\log _2}(1 +\frac{|{h_{ie_j}}{|^2}P_i}{{{N_0}}}),
\end{equation}
where ${\cal E}$ denotes the set of $N$ eavesdroppers. As is shown in [5], the secrecy capacity is given by the difference between the capacity of the main channel and the wiretap channel. Therefore, the secrecy capacity of the $\textrm{U}_i$-BS transmission in the presence of $N$ uncoordinated eavesdroppers is given by
\begin{equation}\label{equa6}
{C^{\textrm{secrecy}}_i} = {C_{ib}} - {C_{ie}}={C_{ib}} -  \mathop {\max }\limits_{e_j \in {\cal{E}}} {C_{{ie_j}}},
\end{equation}
where ${C_{ib}}$ and $C_{i{e_j}}$ are given by (2) and (4), respectively. It is pointed out that all the wireless channels of Fig. 1 are characterized using the Rayleigh fading model. Moreover, the average channel gains of $|h_{ib}|^2$ and $|h_{ie_j}|^2$ are represented by $\sigma_{ib}^2$ and $\sigma_{ie_j}^2$,
respectively.

\section{Multi-user Scheduling Schemes and Secrecy Outage Analysis}
In this section, we propose a multi-user scheduling schemes under the assumption that only the CSIs of the main channels from $M$ users to BS are available without knowing the eavesdroppers' CSIs. We also consider the conventional round-robin scheduling as a benchmark scheme. The closed-form secrecy outage expressions of the round-robin scheduling and the proposed multi-user scheduling are then analyzed over Rayleigh fading channels.

\subsection{Round-Robin Scheduling}
Let us first consider the round-robin scheduling as a baseline. To be specific, the round-robin scheduling allows $M$ users to take turns in transmitting their signals to BS. As is known, when the secrecy capacity falls below a predefined secrecy rate $R_s$, the so-called secrecy outage event is considered to occur in this case. Given that ${\textrm{U}}_i$ transmits its signal to BS, the secrecy outage probability of the ${\textrm{U}}_i$-BS transmission in the presence of $N$ uncoordinated eavesdroppers is expressed as
\begin{equation}\label{equa7}
{P_{out,i}} = \Pr \left( {C_i^{\textrm{secrecy}} < {R_s}} \right),
\end{equation}
where $C_i^{\textrm{secrecy}}$ is given by (6). Substituting $C_i^{\textrm{secrecy}}$ from (6) into (7) yields
\begin{equation}\label{equa8}
\begin{split}
{P_{out,i}}
& = \Pr \left( {\mathop {\max }\limits_{e_j  \in {\cal E}} |h_{ie_j } |^2  > \frac{{|h_{ib} |^2 }}{{2^{R_s } }} - \frac{{(2^{R_s }  - 1)N_0 }}{{2^{R_s } P_i }}} \right),
\end{split}
\end{equation}
which is further obtained as (see Appendix A)
\begin{equation}\label{equa9}
\begin{split}
{P_{out,i}} = 1 - \sum\limits_{n = 0}^{2^N  - 1} {\frac{{( - 1)^{|{\cal E}_n |} }}{{1 + \sum\limits_{e_j  \in {\cal E}_n } {\frac{{\sigma _{ib}^2 }}{{2^{R_s } \sigma _{ie_j }^2 }}} }}\exp ( - \frac{\theta }{{\sigma _{ib}^2 }})},
\end{split}
\end{equation}
where $\theta  = \frac{{(2^{R_s }  - 1)N_0 }}{{P_i }}$, $N$ is the number of eavesdroppers, ${\cal {E}}_n$ represents the $n$-th non-empty subcollection of the elements of ${\cal {E}}$, and $|{\cal {E}}_n|$ is the cardinality of set ${\cal {E}}_n$. In the round-robin scheduling scheme, $M$ users to take turns in transmitting to BS and thus the secrecy outage probability of the conventional round-robin scheduling scheme is given by the mean of $M$ users' secrecy outage probabilities, i.e.,
\begin{equation}\label{equa10}
P_{out}^{\textrm{round}} = \frac{1}{M}\sum\limits_{i = 1}^M {{P_{out,i}}},
\end{equation}
where $M$ is the number of users and ${P_{out,i}}$ is given by (9).

\subsection{Proposed Multi-user Scheduling}
This subsection presents a multi-user scheduling scheme under the assumption that only the CSIs of $M$ users are available without knowing the $N$ eavesdroppers' CSI knowledge. In practice, it is challenging to obtain the eavesdroppers' CSIs, since the eavesdroppers are typically passive. This motivates us to consider the scenario without the eavesdroppers' CSIs. Considering that only the CSIs of users-to-BS channels are known, a user with the highest instantaneous channel gain is selected to transmit to BS, thus the scheduled user is obtained as
\begin{equation}\label{equa11}
\begin{split}
{\textrm{Scheduled User}} =  \arg \mathop {\max }\limits_{i \in {\cal {U}}} {|{h_{ib}}{|^2}},
\end{split}
\end{equation}
where ${\cal {U}}$ represents the set of $M$ users. It can be seen from (11) that only $|h_{ib}|^2$ is assumed in performing the multi-user scheduling without the need of the passive eavesdroppers' CSIs $|h_{ie_j}|^2$. This is attractive from a practical perspective, since the passive eavesdroppers' CSIs are challenging to obtain in practice. For notational convenience, let `$s$' denote the scheduled user by (11). Hence, in the presence of $N$ uncoordinated eavesdroppers, the secrecy capacity of the transmission from the scheduled user ($s$) to BS is given by
\begin{equation}\label{equa12}
C^{\textrm{secrecy}}_s = {C_{sb}} - \mathop {\max }\limits_{{e_j} \in {\cal {E}}} {C_{s{e_j}}},
\end{equation}
where ${C_{sb}}$ and ${C_{s{e_j}}}$ represent the channel capacity from the scheduled user to BS and to eavesdropper ${\textrm{E}}_j$, respectively, which are given by
\begin{equation}\label{equa13}
{C_{sb}} = {\log _2}(1 +\frac{|{h_{sb}}{|^2}P_i}{{{N_0}}}),
\end{equation}
and
\begin{equation}\label{equa14}
{C_{s{e_j}}} = {\log _2}(1 +\frac{|{h_{se_j}}{|^2}P_i}{{{N_0}}}),
\end{equation}
where $|h_{sb}|^2$ and $|h_{s{e_j}}|^2$ represent the fading coefficients of the channels from the scheduled user to BS and to eavesdropper ${\textrm{E}}_j$, respectively. Combining (11)-(14), we can obtain the secrecy outage probability of the proposed multi-user scheduling as
\begin{equation}\label{equa15}
\begin{split}
P_{out}^{\textrm{proposed}} &= \Pr \left( {C_s^{\textrm{secrecy}} < {R_s}} \right) \\
&= \Pr \left( {\mathop {\max }\limits_{e_j  \in {\cal E}} |h_{se_j } |^2  > \frac{{|h_{sb} |^2  - \theta }}{{2^{R_s } }}} \right), \\
 \end{split}
\end{equation}
where $\theta  = \frac{{(2^{R_s }  - 1)N_0 }}{{P_i }}$. Using the law of total probability, we can rewrite (15) as
\begin{equation}\label{equa16}
P_{out}^{\textrm{proposed}}  = \sum\limits_{i = 1}^M {\Pr \left( {\mathop {\max }\limits_{e_j  \in {\cal E}} |h_{ie_j } |^2  > \frac{{|h_{ib} |^2  - \theta }}{{2^{R_s } }},s = i} \right)}.
\end{equation}
Combining (11) and (16), we obtain the secrecy outage probability of the proposed multi-user scheduling as
\begin{equation}\label{equa17}
P_{out}^{\textrm{proposed}}  = \sum\limits_{i = 1}^M {\Pr \left(
\begin{split}
&{\mathop {\max }\limits_{e_j  \in {\cal E}} |h_{ie_j } |^2  > \frac{{|h_{ib} |^2  - \theta }}{{2^{R_s } }}},\\
&\mathop {\max }\limits_{\scriptstyle k \in {\cal U} \hfill \atop\scriptstyle k \ne i \hfill} |h_{kb} |^2  < |h_{ib} |^2
\end{split}
\right)}.
\end{equation}
Proceeding as in Appendix B, we obtain (18) at the top of the following page,
\begin{figure*}
\begin{equation}\label{equa18}
\begin{split}
P_{out}^{\textrm{proposed}}  =& \sum\limits_{i = 1}^M {\sum\limits_{m = 0}^{2^{M - 1}  - 1} {\frac{{( - 1)^{|{\cal U}_m |} }}{{1 + \sum\limits_{k \in {\cal U}_m } {\frac{{\sigma _{ib}^2 }}{{\sigma _{kb}^2 }}} }}\left( {1 - \exp ( - \sum\limits_{k \in {\cal U}_m } {\frac{\theta }{{\sigma _{kb}^2 }}}  - \frac{\theta }{{\sigma _{ib}^2 }})} \right)} }  \\
&+ \sum\limits_{i = 1}^M {\sum\limits_{m = 0}^{2^{M - 1}  - 1} {\sum\limits_{n = 1}^{2^N  - 1} {\frac{{( - 1)^{|{\cal U}_m | + |{\cal E}_n | + 1} }}{{1 + \sum\limits_{e_j  \in {\cal E}_n } {\frac{{\sigma _{ib}^2 }}{{2^{R_s } \sigma _{ie_j }^2 }}}  + \sum\limits_{k \in {\cal U}_m } {\frac{{\sigma _{ib}^2 }}{{\sigma _{kb}^2 }}} }}\exp ( - \sum\limits_{k \in {\cal U}_m } {\frac{\theta }{{\sigma _{kb}^2 }}}  - \frac{\theta }{{\sigma _{ib}^2 }})} } }  \\
 \end{split}
\end{equation}
\end{figure*}
where ${\cal U}_0$ represents an empty set, ${\cal {U}}_m$ represents the $m$-th non-empty subset of ${\cal {U}}-\{{{\textrm{U}}_i}\}$, and`$-$' is the set difference.

\section{Numerical Results and Discussions}
In this section, we present numerical secrecy outage probability results of the conventional round-robin scheduling and the proposed multi-user scheduling schemes.
For notational convenience, $\lambda_{me}$ is used to represent the ratio of $\sigma^2_{ib}$ to $\sigma^2_{ie_j}$, which is called the main-to-eavesdropper ratio (MER) throughout this paper. Moreover, let $\gamma$ denote the ratio of $P_i$ to $N_0$, i.e., $\gamma=P_i/N_0$. Fig. 2 plots (10) and (18) as a function of the MER $\lambda_{me}$ for different number of eavesdroppers $N$. The simulated secrecy outage probability results for the conventional round-robin scheduling and the proposed multi-user scheduling schemes are also given in this figure. One can see from Fig. 2 that as the number of eavesdroppers increases from $N=2$ to $8$, the secrecy outage probabilities of the round-robin scheduling and the proposed multi-user scheduling schemes both increase. This means that with an increasing number of eavesdroppers, it would be more likely to succeed in intercepting the wireless transmission. It is also shown from Fig. 2 that for both the cases of $N=2$ and $N=8$, the proposed multi-user scheduling outperforms the conventional round-robin scheduling in terms of the secrecy outage probability. Additionally, the theoretical secrecy outage probabilities match the simulated results well.

\begin{figure}
  \centering
  {\includegraphics[scale=0.52]{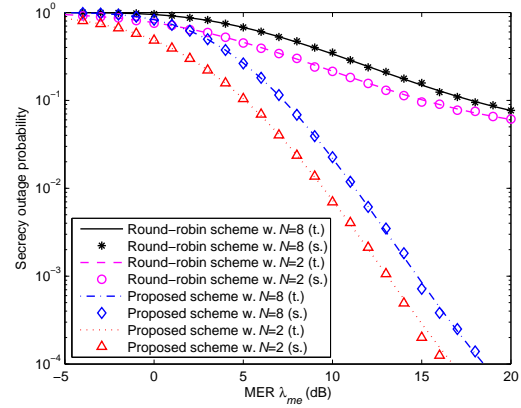}\\
  \caption{Secrecy outage probability versus MER $\lambda_{me}$ of the round-robin scheduling and the proposed multi-user scheduling schemes for different number of eavesdroppers $N$ with $M=4$, $R_s=0.5{\textrm{ bit/s/Hz}}$, and $\gamma=10{\textrm{ dB}}$.}\label{Fig2}}
\end{figure}

Fig. 3 illustrates the secrecy outage probability versus MER $\lambda_{me}$ of the round-robin scheduling and the proposed multi-user scheduling schemes for different number of users $M$ with $N=4$, $R_s=0.5{\textrm{  bit/s/Hz}}$, and $\gamma=10{\textrm{ dB}}$. As shown in Fig. 3, as the number of users increases from $M=4$ to $8$, the secrecy outage probability of the conventional round-robin scheduling remains unchanged. By contrast, the secrecy outage performance of the proposed multi-user scheduling scheme corresponding to $M=8$ is significantly better than that corresponding to $M=4$. This further confirms the security advantage of the proposed multi-user scheduling over the conventional round-robin scheduling.
\begin{figure}
  \centering
  {\includegraphics[scale=0.52]{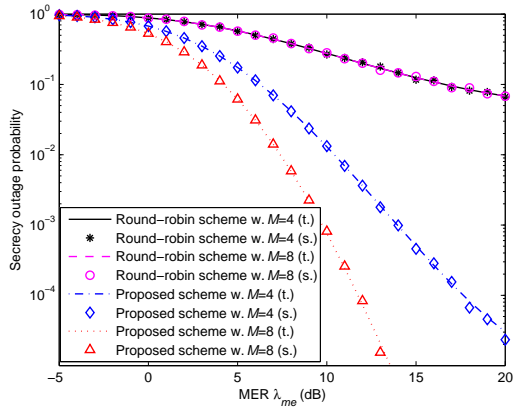}\\
  \caption{Secrecy outage probability versus MER $\lambda_{me}$ of the round-robin scheduling and the proposed multi-user scheduling schemes for different number of users $M$ with $N=4$, $R_s=0.5{\textrm{ bit/s/Hz}}$, and $\gamma=10{\textrm{ dB}}$.}\label{Fig3}}
\end{figure}

In Fig. 4, we show the secrecy outage probability versus the number of users $M$ of the round-robin scheduling and the proposed multi-user scheduling schemes for different secrecy rates with $N=4$ and $\lambda_{me}=\gamma=10{\textrm{ dB}}$. It is observed from Fig. 4 that the secrecy outage probability of the conventional round-robin scheduling is insusceptible to the number of users. By contrast, the secrecy performance of the proposed multi-user scheduling scheme is improved as the number of users increases. Fig. 4 also shows that for both the cases of $R_s=2{\textrm{ bit/s/Hz}}$ and $R_s=1{\textrm{ bit/s/Hz}}$, the proposed multi-user scheduling performs better than the conventional round-robin scheduling in terms of the secrecy outage probability, especially with a large number of users.
\begin{figure}
  \centering
  {\includegraphics[scale=0.52]{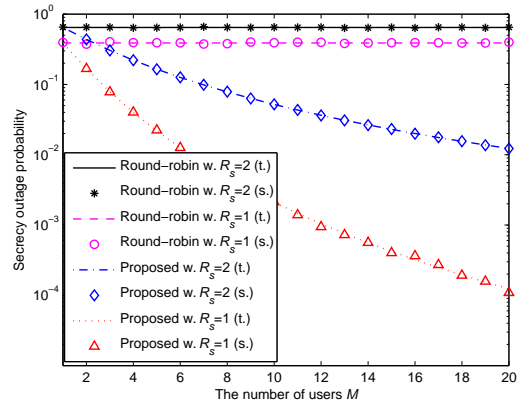}\\
  \caption{Secrecy outage probability versus the number of users $M$ of the round-robin scheduling and the proposed multi-user scheduling schemes for different secrecy rates with $N=4$ and $\lambda_{me}=\gamma=10{\textrm{ dB}}$.}\label{Fig4}}
\end{figure}

\section{Conclusion}
In this paper, we studied the secrecy outage performance of a multi-user multi-eavesdropper wireless system, where multiple users are intended to transmit to BS in the presence of multiple eavesdroppers attempting to tap the users-to-BS transmissions. We proposed a multi-user scheduling scheme that only needs the CSIs of the users-to-BS channels without knowing the passive eavesdroppers' CSIs. We also considered the conventional round-robin scheduling as a benchmark. Closed-form secrecy outage probability expressions of the round-robin scheduling and the proposed multi-user scheduling were derived over Rayleigh fading channels. Numerical performance comparisons between the conventional round-robin scheduling and the proposed multi-user scheduling schemes were carried out in terms of the secrecy outage probability. It was showed that the secrecy outage performance of the proposed multi-user scheduling is significantly better than that of the conventional round-robin scheduling. Additionally, upon increasing the number of users, the secrecy outage probability of the proposed multi-user scheduling scheme decreases, showing the security advantage of exploiting the multi-user scheduling against eavesdropping attacks.

\appendices
\section{Proof of (9)}
For notational convenience, let $X$ and $Y_j$ denote $|h_{ib} |^2$ and $|h_{ie_j } |^2$, respectively. Noting that $|h_{ib}|^2$ and $|h_{ie_j}|^2$ are independent and exponentially distributed random variables with respective means of $\sigma^2_{ib}$ and $\sigma^2_{ie_j}$, we can rewrite (8) as
\begin{equation}
\begin{split}
 P_{out,i}  &= \Pr \left( {\mathop {\max }\limits_{e_j  \in {\cal E}} Y_j  > \frac{X}{{2^{R_s } }} - \frac{\theta }{{2^{R_s } }}} \right) \\
&= 1 - \int_{\theta}^\infty  {\prod\limits_{e_j  \in {\cal E}} {\left( {1 - \exp ( - \frac{{x - \theta }}{{2^{R_s } \sigma _{ie_j }^2 }})} \right)} }  \\
&\quad\quad\quad\quad\quad \times \frac{1}{{\sigma _{ib}^2 }}\exp ( - \frac{x}{{\sigma _{ib}^2 }})dx ,
 \end{split} \tag{A.1}\label{A.1}
\end{equation}
where $\theta  = \frac{{(2^{R_s }  - 1)N_0 }}{{P_i }}$. By using the binomial theorem, the term ${\prod\limits_{e_j  \in {\cal E}} {\left( {1 - \exp ( - \frac{{x - \theta }}{{2^{R_s } \sigma _{ie_j }^2 }})} \right)} }$ can be expanded as
\begin{equation}
\begin{split}
&\prod\limits_{e_j  \in {\cal E}} {\left( {1 - \exp ( - \frac{{x - \theta }}{{2^{R_s } \sigma _{ie_j }^2 }})} \right)} \\
&= \sum\limits_{n = 0}^{2^N  - 1} {( - 1)^{|{\cal E}_n |} \exp ( - \sum\limits_{e_j  \in {\cal E}_n } {\frac{{x - \theta }}{{2^{R_s } \sigma _{ie_j }^2 }}} )},
\end{split}\tag{A.2}\label{A.2}
\end{equation}
where $\sum\limits_{e_j  \in {\cal E}_0 } {\frac{1}{{2^{R_s } \sigma _{ie_j }^2 }}}  = 0$, ${\cal {E}}_0$ is an empty set, ${\cal {E}}_n$ represents the $n$-th non-empty subset of ${\cal {E}}$, and $|{\cal {E}}_n|$ is the cardinality of ${\cal {E}}_n$. Substituting (A.2) into (A.1) yields
\begin{equation}
\begin{split}
P_{out,i} & = 1 - \sum\limits_{n = 0}^{2^N  - 1} {\int_\theta ^\infty  {\frac{{( - 1)^{|{\cal E}_n |} }}{{\sigma _{ib}^2 }}\exp ( - \frac{x}{{\sigma _{ib}^2 }} - \sum\limits_{e_j  \in {\cal E}_n } {\frac{{x - \theta }}{{2^{R_s } \sigma _{ie_j }^2 }}} )dx} }  \\
&= 1 - \sum\limits_{n = 0}^{2^N  - 1} {\frac{{( - 1)^{|{\cal E}_n |} }}{{1 + \sum\limits_{e_j  \in {\cal E}_n } {\frac{{\sigma _{ib}^2 }}{{2^{R_s } \sigma _{ie_j }^2 }}} }}\exp ( - \frac{\theta }{{\sigma _{ib}^2 }})},
\end{split}\tag{A.3}\label{A.3}
\end{equation}
which is (9).

\section{Derivation of (18)}
Denoting $|h_{ib} |^2  = X$, $|h_{ie_j } |^2  = Y_j$, and $|h_{kb} |^2  = Z_k$, we can rewrite (17) as
\begin{equation}
P_{out}^{\textrm{proposed}}  = \sum\limits_{i = 1}^M {\Pr \left( {\mathop {\max }\limits_{e_j  \in {\cal E}} Y_j  > \frac{{X - \theta }}{{2^{R_s } }},\mathop {\max }\limits_{\scriptstyle k \in {\cal U} \hfill \atop
\scriptstyle k \ne i \hfill} Z_k  < X} \right)}.\tag{B.1}\label{B.1}
\end{equation}
Noting that random variables $|h_{ib} |^2 $, $|h_{ie_j } |^2 $ and $|h_{kb} |^2 $ are independent and exponentially distributed with respective means of $\sigma^2_{ib}$, $\sigma^2_{ie_j}$ and $\sigma^2_{kb}$, we obtain
\begin{equation}
\begin{split}
P_{out}^{\textrm{proposed}}  &= \sum\limits_{i = 1}^M {\int_0^\theta  {\prod\limits_{\scriptstyle k = 1 \hfill \atop
\scriptstyle k \ne i \hfill}^M {\left( {1 - \exp ( - \frac{x}{{\sigma _{kb}^2 }})} \right)} \frac{1}{{\sigma _{ib}^2 }}\exp ( - \frac{x}{{\sigma _{ib}^2 }})dx} }  \\
&+ \sum\limits_{i = 1}^M {\int_\theta ^\infty  {\left[ {1 - \prod\limits_{j = 1}^N {\left( {1 - \exp ( - \frac{{x - \theta }}{{2^{R_s } \sigma _{ie_j }^2 }})} \right)} } \right]} }  \\
&\quad\quad \times \prod\limits_{\scriptstyle k = 1 \hfill \atop
\scriptstyle k \ne i \hfill}^M {\left( {1 - \exp ( - \frac{x}{{\sigma _{kb}^2 }})} \right)} \frac{1}{{\sigma _{ib}^2 }}\exp ( - \frac{x}{{\sigma _{ib}^2 }})dx. \\
\end{split}\tag{B.2}\label{B.2}
\end{equation}
Using the binomial theorem, we have
\begin{equation}
\begin{split}
&\prod\limits_{\scriptstyle k = 1 \hfill \atop
\scriptstyle k \ne i \hfill}^M {\left( {1 - \exp ( - \frac{x}{{\sigma _{kb}^2 }})} \right)}  \\
&= \sum\limits_{m = 0}^{2^{M - 1}  - 1} {( - 1)^{|{\cal U}_m |} \exp ( - \sum\limits_{k \in {\cal U}_m } {\frac{x}{{\sigma _{kb}^2 }}} )},
\end{split}\tag{B.3}\label{B.3}
\end{equation}
where $\sum\limits_{k \in {\cal U}_0 } {\frac{1}{{\sigma _{kb}^2 }}}  = 0$, ${\cal U}_0$ represents an empty set, ${\cal {U}}_m$ represents the $m$-th non-empty subset of ${\cal {U}}-\{{{\textrm{U}}_i}\}$, and `$-$' is the set difference. Similarly, term $\prod\limits_{j = 1}^N {\left( {1 - \exp ( - \frac{{x - \theta }}{{2^{R_s } \sigma _{ie_j }^2 }})} \right)}$ can be expanded as
\begin{equation}
\begin{split}
&\prod\limits_{j = 1}^N {\left( {1 - \exp ( - \frac{{x - \theta }}{{2^{R_s } \sigma _{ie_j }^2 }})} \right)}  \\
&= 1 + \sum\limits_{n = 1}^{2^N  - 1} {( - 1)^{|{\cal E}_n |} \exp ( - \sum\limits_{e_j  \in {\cal E}_n } {\frac{{x - \theta }}{{2^{R_s } \sigma _{ie_j }^2 }}} )},
\end{split}\tag{B.4}\label{B.4}
\end{equation}
where ${\cal {E}}_n$ represents the $n$-th non-empty subset of ${\cal {E}}$. Substituting (B.3) and (B.4) into (B.2) and performing the integration, we can arrive at (18).

\clearpage

\end{document}